\documentclass[a4paper,11pt]{article}
\usepackage{pos}
\usepackage{booktabs}
\usepackage[capitalize]{cleveref}
\usepackage{bm}
\usepackage[utf8]{inputenc} 

\newcommand{\bra}[1]{\langle #1 |} 
\newcommand{\ket}[1] {| #1 \rangle} 
\newcommand{\matrixel}[3]{\langle #1 | #2 | #3 \rangle} 
\newcommand{\vacuum}[0]{\Omega}

\newcommand\underrel[3][]{\mathrel{\mathop{#3}\limits_{%
      \ifx c#1\relax\mathclap{#2}\else#2\fi}}}

\newcommand\st[1]{} 

\newcommand\citethis[1]{\text{\color{red}CITE!!}}
\definecolor{darkgreen}{rgb}{0.0, 0.545098, 0.0}

\title{Toward inclusive observables with staggered quarks: the smeared $R$~ratio}
\ShortTitle{Toward inclusive observables with staggered quarks}

\author[a]{Thomas Blum}
\author[b,c]{William I. Jay}
\author[a]{Luchang Jin}
\author[d]{Andreas S. Kronfeld}
\author*[a]{Douglas B. A. Stewart}

\affiliation[a]{Department of Physics, University of Connecticut, Storrs, CT 06269-3046, USA}

\affiliation[b]{Center for Theoretical Physics, Massachusetts Institute of Technology, Cambridge, MA 02139, USA}

\affiliation[c]{Department of Physics, Colorado State University, Fort Collins, CO 80523, USA}

\affiliation[d]{Theory Division, Fermi National Accelerator Laboratory, Batavia, Illinois, 60510, USA}

\emailAdd{thomas.blum@uconn.edu}
\emailAdd{william.jay@colostate.edu}
\emailAdd{luchang.jin@uconn.edu}
\emailAdd{ask@fnal.gov}
\emailAdd{douglas.stewart@uconn.edu}

\abstract{
Inclusive hadronic observables are ubiquitous in particle and nuclear physics.
Computation of these observables using lattice QCD is challenging due the presence of a difficult inverse problem.
As a stepping stone to more complicated observables, we report on progress to compute the smeared $R$~ratio with staggered quarks using the spectral reconstruction algorithm of Hansen, Lupo, and Tantalo.
We compare staggered-quark results on two ensembles to domain-wall results on a single ensemble and to the Bernecker-Meyer parameterization.
This work utilizes two ensembles generated by the MILC collaboration using highly improved staggered quarks and one ensemble generated by the RBC/UKQCD collaboration using domain-wall quarks.
Possible strategies for controlling opposite-parity effects associated with staggered quarks are discussed.}

\FullConference{The 41st International Symposium on Lattice Field Theory (LATTICE2024)\\
 28 July - 3 August 2024\\
Liverpool, UK\\}

\begin{document}
\maketitle

\section{Introduction}\label{sec:intro}
    Inclusive scattering processes are experimentally important throughout particle physics.
    For instance, deep inelastic scattering measurements  ($ep \to eX$) were crucial in establishing QCD as the theory of the strong interactions.
    Inclusive weak decays of hadrons (like $B\to X\ell\nu$) play an important role in determining the elements of the CKM matrix~\cite{shoji:2017Bmeson,gambino:2020semilep,gambino:2022heavy}.
    Upcoming accelerator-based neutrino experiments like DUNE will be sensitive to the inclusive vector and axial structure functions of the nucleon (\emph{e.g.}, $\nu_\ell p \to \ell X$)~\cite{fukaya:2020lN}.
    The hadronic contribution to the anomalous magnetic moment ($g-2$) of the muon is closely related to measurements of $e^+e^- \to {\rm hadrons}$.
    The common field-theoretic object in these processes is the hadronic tensor $W^{\mu\nu}$ which encodes the hadronic response of a particular state, be it the vacuum or a particular hadron, to external currents.
    
    The hadronic tensor is proportional to the differential cross section, $d\sigma \propto W_{\mu\nu} L^{\mu\nu}$, where the leptonic tensor $L^{\mu\nu}$ is taken to be calculable perturbatively
    (see, e.g., Ref.~\cite{Manohar:1992tz}).
    The hadronic tensor is defined for a given hadronic state $|H\rangle$ with four-momentum $p = (p^0, \bm{p})$ and external current $J^\mu$ via the QCD correlation function: 
    $W_{\mu\nu} \propto \int d^4x\, e^{iq\cdot x} \langle H, \bm{p}|J_{\mu}(x)J_{\nu}(0)|H,\bm{p}\rangle$.
    In the low-energy, non-perturbative regime, it is natural to calculate the Euclidean version of this quantity using lattice QCD.
    The Euclidean hadronic tensor is related to the Minkowski quantity via the Laplace transform,
    \begin{equation} 
        W^{\text{Euc}}_{\mu\nu}(\bm{q}^2, \tau) = \int d\omega\, e^{-\omega\tau} W_{\mu\nu}(\bm{q}^2, \omega),
        \label{eq:laplace}
    \end{equation}
    where $\tau$ is the Euclidean time and $\omega$ is energy.
    Inverting this equation presents an delicate numerical challenge, for which various numerical methods have been proposed in the literature~\cite{Pijpers:1992,Hansen:2019idp,PhysRevB.96.035147,PhysRevE.95.061302,Itou:2020azb,doi:10.7566/JPSJ.89.012001,PhysRevB.101.035144,PhysRevB.98.035104,Shi:2022yqw,Chen:2021giw,Kades:2019wtd,Zhou:2023pti,Lechien:2022ieg,Wang:2021jou,10.1093/gji/ggz520,DelDebbio:2021whr,Candido:2023nnb,Horak:2021syv,Pawlowski:2022zhh,Horak:2023xfb,Rothkopf:2022fyo,Rothkopf:2022ctl,Bulava:2023mjc,Bergamaschi:2023xzx,shoji}.
    
    This proceedings reports progress on a pilot study to compute the hadronic tensor for inclusive electromagnetic scattering of the pion ($e\pi \to e X$) using lattice QCD.
    While the hadronic tensor is the eventual goal, the $R$~ratio is a useful intermediate milestone observable.
    It is defined as the ratio of the cross sections for $e^+e^-\to\text{ hadrons}$ and $e^+e^-\to\mu^+\mu^-$,
    $R(s) \equiv \tfrac{\sigma(e^+e^- \to \text{hadrons})}{\sigma(e^+e^- \to \mu^+\mu^-)}$,
    and is related to the Euclidean vector-vector correlator by~\cite{Lehner:2020hvp}
    \begin{align}
    C(\tau)
    = \sum_{\mu=1}^3 \int d^3\bm{x}\, 
        \bra{0}J_\mu^\text{EM}(\tau,\bm{0})J_\mu^\text{EM}(0,\bm{0})\ket{0} 
    = \frac{1}{12\pi^2}\int_0^\infty d\left(\sqrt{s}\right)\,
        R(s) s e^{-\sqrt{s}\tau}.
    \end{align}
    The $R$~ratio is closely related to calculations of the hadronic contributions to
    the anomalous magnetic moment of the muon~\cite{Aoyama:2020ynm}.
    We shall use a smeared version of $R$ to develop our analysis pipeline.
    
    The smeared $R$~ratio has previously been computed using the Hansen, Lupo, Tantalo (HLT) method~\cite{Hansen:2019idp} by 
    the ETM Collaboration using twisted-mass fermions~\cite{DeSantis:2022qht,ExtendedTwistedMassCollaborationETMC:2022sta}.
    In comparison, we compute the smeared $R$~ratio using staggered fermions.
    Staggered correlation functions contain states with both positive and negative parity~\cite{DeGrand:2006zz}.
    We are exploring strategies to extract smeared spectral densities of definite parity using staggered fermions.
    For comparison, we also analyze the spectral density computed using domain-wall fermions (DWF).

\section{Spectral Reconstruction}\label{sec:methods}
    Calculation of a spectral function from a Euclidean-time correlator amounts to a challenging problem in numerical analytic continuation, a perspective recently emphasized in, \emph{e.g.}, Refs.~\cite{Bergamaschi:2023xzx,Huang:2022qsb,PhysRevLett.126.056402}.
    In general regulators such as smearing are used to improve the stability of inverse problems.
    
    Additionally, computations are performed on a lattice with finite volume, which provides an IR cutoff for the theory and renders the spectrum discrete.
    A generic Euclidean correlator encodes the finite-volume spectral function consisting of a discrete sum of delta functions, whose density increases as a function of volume and energy.
    Connecting finite-volume results with their infinite-volume counter parts also motivates the use of smearing.
    The spectral function is defined as the result of an ordered limit $\rho(\omega) = \lim_{\sigma\to 0}\lim_{L\to\infty}\tilde{\rho}_L(\omega)$,
    where $\tilde{\rho}_L(\omega)$ is the smeared finite-volume spectral function~\cite{Hansen:2017mnd}.

\subsection{The Reconstruction Algorithm of Hansen, Lupo, and Tantalo}\label{sec:HLT}
    The following subsection is a brief summary of Ref.~\cite{Hansen:2019idp}.
    A general finite-volume correlation function $C(\tau)$, where $\tau$ is Euclidian time, may be written as the Laplace transform of a finite-volume spectral function $\rho_L(\omega)$:
    \begin{equation}
        C(\tau) = \int^\infty_0 d\omega e^{-\omega\tau} \rho_L(\omega).
        \label{eq:corr_transform}
    \end{equation}
    A smeared spectral function may be defined by convolution with a smearing function
    $\Delta(\omega^*,\omega)$ constructed from a linear combination of suitably chosen basis functions $b(\omega, \tau)$:
    \begin{equation}
        \Delta(\omega^*,\omega) = \sum_{\tau=1}^{\tau_\text{max}} g_\tau(\omega^*) b(\omega,\tau).
        \label{eq:BGsmear}
    \end{equation}
    For a lattice with finite temporal extent, the basis functions take the usual form, 
    \begin{align}
    b(\omega,\tau) = e^{-\omega\tau}+e^{-\omega(T-\tau)}. \label{eq:basis_functions} 
    \end{align}
    In the limit of very large temporal extent, these simplify, $b(\omega,\tau) \stackrel{T\to\infty}{=} e^{-\omega\tau}$.
    From \cref{eq:corr_transform,eq:BGsmear}, the smeared finite-volume spectral function $\tilde{\rho}_L(\omega^*)$ takes the form:
    \begin{align}
        \tilde{\rho}_L(\omega^*)&=\int_0^\infty d\omega\rho_L(\omega)\Delta(\omega^*,\omega)=\int_0^\infty d\omega\rho_L(\omega)\sum_{\tau=1}^{\tau_\text{max}} g_\tau(\omega^*) b(\omega,\tau) \nonumber\\
        &=\sum_{\tau=1}^{\tau_\text{max}} g_\tau(\omega^*)\int_0^\infty d\omega\rho_L(\omega)e^{-\omega(\tau)} 
        =\sum_{\tau=1}^{\tau_\text{max}} g_\tau(\omega^*)C(\tau) \label{eq:BG_smeared_rho}.
    \end{align}
    The HLT method defines the coefficients $\mathbf{g(\omega^*}) \equiv (g_0(\omega^*),\ldots,g_{\tau_{\text{max}}}(\omega^*))$ in \cref{eq:BGsmear}
    as the solution to a convex optimization problem, namely, minimization of the functional
    $W[\mathbf{g}] = \frac{A[\mathbf{g}]}{A[\mathbf{0}]} + \lambda B[\mathbf{g}]$,
    \begin{align}
         A[\mathbf{g}] &= \int_{\omega_0}^\infty d\omega w_\alpha(\omega) \bigg| \Delta_\sigma^{\text{in}}(\omega^*,\omega) - \sum_{\tau=1}^{\tau_\text{max}} g_\tau(\omega^*) b(\omega,\tau) \bigg|^2
        \label{eq:A_func}\\
         B[\mathbf{g}] &= B_\text{norm} \sum_{\tau_1,\tau_2 = 1}^{\tau_\text{max}} g_{\tau_1} g_{\tau_2} \text{Cov}(\tau_1,\tau_2),\;\; B_\text{norm} = \frac{1}{C(a)^2},
         \label{eq:B_func}
    \end{align}
    where the systematic functional in \cref{eq:A_func} is the weighted $\ell^2$-norm between the input and output smearing functions, the statistical functional in \cref{eq:B_func} takes errors in the input data into account during the reconstruction, and $\lambda$ is a hyperparameter that is tuned to reduce the systematic error of the final spectral reconstruction.
    For the reconstructions shown in this proceedings, a normalized Gaussian smearing function is used,
    \begin{equation}\label{eq:gaussian}
        \Delta^{\text{in}}_\sigma(\omega^*,\omega) = \frac{1}{\sqrt{2\pi}\sigma} e^{-\frac{(\omega-\omega^*)^2}{2\sigma^2}},
    \end{equation}
    and statistical and systematic errors are added in quadrature following the stability analysis defined in Ref.~\cite{ExtendedTwistedMassCollaborationETMC:2022sta}.

\section{Lattice QCD Calculation}\label{sec:lattices}

    \begin{table}[t]
        \centering

        \begin{tabular}{c c c c c c c}
            \hline\hline
            Fermion discretization & $\approx a~{\rm [fm]}$ & $N_s^3 \times N_t$  & $\approx M_{\pi}~{\rm [MeV]}$ & $N_{\rm configs}$ & $N_{\rm low}$ & $N_{\rm high}$ \\
            \hline
            Staggered & 0.15 & $32^3 \times 48$  & 138 & 37 & 4000 & 144 \\
            Staggered & 0.12 & $48^3 \times 64$  & 135 & 98 & 4000 & 192 \\
            Domain Wall & 0.114 & $48^3 \times 96$ & 139 & 112 & - & - \\
            \hline\hline
        \end{tabular}
        \caption{
        Summary of gauge-field ensembles used in this calculation.
        $M_\pi$ refers to the approximate value of the pion mass in the sea; for the staggered-quark ensembles, $M_\pi$ refers to the Goldstone pion with pseudoscalar taste.
        The parameter $N_{\rm low}$ refers to the number of eigenmodes of the Dirac operator; for staggered-quark ensembles these were computed exactly for all-to-all propagators.
        For the domain-wall ensemble, 2048 point-source propagators were used instead of all-to-all propagators.
        \label{tab:ensembles}
        }
    \end{table}

    The present work analyzes correlations functions computed on three different gauge-field ensembles, summarized in \cref{tab:ensembles}.
    The two staggered-quark ensembles were generated by the MILC collaboration using $2+1+1$ dynamical flavors of quarks with physical masses using a one-loop Symanzik-improved gauge action and the highly improved staggered quark (HISQ) action~\cite{MILC:2012znn}.
    For the valence quarks, a fat-link-only variant of the HISQ action is used which omits the Naik term~\cite{Aubin:2022hgm}.
    Both staggered-quark ensembles have similar physical volumes.
    The domain-wall ensemble is the $48\text{I}$ ensemble generated by the RBC/UKQCD collaboration~\cite{RBC:2014ntl,RBC:2023pvn}.
    The physical volume $V \approx (5-5.75~{\rm fm})^3$ is similar for all three ensembles.
    Vector-current correlation functions are defined with $N_f=2$ (light only) and $N_f=2+1$ (light and strange) flavors of valence quarks near their physical values for the staggered and domain-wall ensembles, respectively.
    Staggered correlation functions are computed using all-to-all (A2A) methods for the the quark propagators~\cite{Foley:2005ac} using \texttt{Grid}~\cite{Boyle:2016lbp} and the measurement workflow management system \texttt{Hadrons}~\cite{antonin_portelli_2023_8023716}.
    On each configuration, a total of $N_{\rm low}$ eigenmodes of the Dirac operator are computed exactly (see
    \cref{tab:ensembles}).
    Domain-wall correlation functions are computed as in Ref.~\cite{RBC:2023pvn}.

\section{The Staggered Spectrum}\label{sec:staggered}
    The spectral decomposition for a generic staggered-quark correlation functions takes the form
    \begin{align}
    C(\tau)
        &= \sum_{n=0}^\infty (-1)^{n(\tau+1)} \frac{\left|\matrixel{\vacuum}{\mathcal{O}}{n}\right|^2}{2E_n }
        \left( e^{-E_n \tau} + e^{-E_n(N_\tau-\tau)} \right) \label{eq:staggered_spectral_decomp}
    \end{align}
    and includes contributions from states with both positive and negative parity.
    For the vector-current correlators of interest, these states have quantum numbers $J^P = 1^-$ or $1^+$.
    In infinite volume, the former include resonances like the $\rho$, while the latter includes resonances like the $b_1$.
    States with $J^P = 1^-$ decay in Eucildean time, while opposite-parity states with $J^P = 1^+$ decay and oscillate due to the factor $(-1)^{\tau+1}$.
    The spectral decomposition for domain-wall correlators is simpler, involving states of only a single parity and no oscillating contributions.

    The remainder of this section discusses possible strategies for handling the opposite-parity contributions in smeared spectral reconstructions using staggered correlation functions.
    
\subsection{Removing the Effects of Oscillating States}\label{sec:OSS}
    In principle, \cref{eq:staggered_spectral_decomp} can be understood as defining separate spectral densities $\rho_{\rm even}$ and $\rho_{\rm odd}$ associated with the even and odd time-slices, respectively.
    Diagonal correlation functions (\emph{i.e.}, ones with the same source and sink operators) 
    generically contain both positive and negative spectral weights:
    $\rho_{\rm{even}} = \rho_- + \rho_+$, and $\rho_{\rm{odd}} = \rho_- - \rho_+$,
    where $\rho_\pm$ is the spectral density associated with states of parity $P=\pm1$.
    By taking linear combinations of these mixed-parity spectra, we arrive at the spectral functions of definite parity,
    \begin{equation}\label{eq:rho_pm}
        \rho_\mp = \frac{1}{2}(\rho_{\rm{even}} \pm \rho_{\rm{odd}}).
    \end{equation}
    These observations suggests an analysis approach which applies the HLT algorithm (or any method for extracting a smeared spectral function) to the even and odd time-slices separately.
    A final post processing step using \cref{eq:rho_pm} would then yield the desired spectral function.
    We adopt this strategy for the reconstructions in this proceedings.

    From a practical perspective, the success of this approach will depend on the statistical precision of a given input correlator.
    All else being equal, a reconstruction method is likely to perform best on precise data with many points.
    Computing $\rho_{\rm even}$ and $\rho_{\rm odd}$ requires eliminating half of the available data in the individual reconstructions, which may prove prohibitive depending on the statistical quality and Euclidean time extent of a given dataset.
    With this in mind, it is worthwhile to consider alternative analysis approaches for staggered quarks.
    Note that the following ideas are in a developmental stage not yet suitable for inclusion in this proceedings and will be explored more thoroughly in a forthcoming publication.
    
\subsubsection{Oscillating-state Subtraction}\label{sec:corr_sub}
    In many physical contexts, contributions from opposite-parity states are simply nuisance parameters which must be controlled in the approach to the continuum and infinite-volume limits.
    The form of the spectral decomposition in \cref{eq:staggered_spectral_decomp} suggests writing a generic staggered correlator in the form
    $C(\tau) = C^{\rm{decay}}(\tau) + C^{\rm{osc}}(\tau)$.
    For large Euclidean times, contributions from excited states are exponentially suppressed, and $C(\tau)$ is dominated by low-lying states.
    If the dominant effects of the low-lying oscillating states can be subtracted explicitly, 
    \begin{align}
    C^{\rm{decay}}(\tau) = C(\tau) - C^{\rm{osc}}(\tau),
    \label{eq:subtraction}
    \end{align}
    it may be possible to reconstruct the desired smeared spectral function directly using all time-slices.
    To this end, we consider a two-stage process.
    First, a multi-exponential fit (\emph{e.g.}, with \texttt{corrfitter}~\cite{corrfitter}) is used to model the behavior of the low-lying states, which gives a parametric form for $C^{\rm osc}(\tau)$ appearing in \cref{eq:subtraction}.
    Second, the spectral reconstruction algorithm is applied to the to the decaying part of the correlator after subtracting off the parametric description of $C^{\rm osc}(\tau)$.
    Of course, the multi-state fit amounts on some level to a spectral reconstruction.
    If the aim is to determine the low-lying spectrum of a correlator, one must carefully scrutinize the added utility of spectral reconstruction algorithmsm like HLT.
    Important work in this direction has been carried out in Ref.~\cite{Bennett:2024cqv}, which compares spectral reconstructions to more traditional spectroscopy methods.    

\subsubsection{Correlator Interpolation}\label{sec:interpolation}
    Taking linear combinations at the level of spectral functions via \cref{eq:rho_pm} reduces undesirably the number of available time-slices.
    Instead, one can imagine taking linear combinations directly at the level of the correlator to increase the number of time-slices available for spectral reconstruction.
    This line of thinking motivate interpolating the correlator separately on the even and odd time-slices.
    Let $\tilde{C}(\tau)$ denote an interpolating function for the data $C(\tau)$ and similarly for $\tilde{C}^{\rm even}(\tau)$ and $\tilde{C}^{\rm odd}(\tau)$.
    Then interpolating functions coupling to states of definite parity are given by
    \begin{align}
        C(\tau)
            &\to \tilde{C}^{\pm}(\tau) = \frac{1}{2} \Big( \tilde{C}^{\rm{even}}(\tau) \pm \tilde{C}^{\rm{odd}}(\tau) \Big)\label{eq:corr_interpolator_pm}.
    \end{align}
    Similar use of interpolating functions has been considered in the context of the muon ($g-2$), \emph{e.g.}, in Refs.~\cite{Lahert:2023ore,Lahert:2024vvu}.
    Substituting \cref{eq:corr_interpolator_pm} into \eqref{eq:BG_smeared_rho} then yields spectral functions $\rho_\pm$ with definite parity.
    Spline interpolants (perhaps computed from $\log C(\tau)$, which is slowly varying) are readily calculated with existing software packages, including \texttt{gvar}~\cite{gvar}.

\subsubsection{Changing the Basis}\label{sec:stag_basis}
    The choices of input smearing function $\Delta_\sigma^{\text{in}}(\omega^*, \omega)$ and basis functions $b(\omega, \tau)$
    are largely arbitrary up to the requirements that $\Delta^{\text{in}}_\sigma(\omega^*,\omega) \to \delta(\omega-\omega^*)$ as $\sigma \to 0$.
    Practicality dictates that reasonable choices should give functions that can be implemented numerically, \emph{e.g.}, with quickly convergent sums and integrals appearing in the intermediate numerical steps in the HLT algorithm.
    One possibility to handle the staggered spectrum in one fell swoop would be to modify the basis functions to include the characteristic factors of $(-1)^{n(\tau + 1)}$ directly.
    Such a modification, while theoretically capable of producing the staggered spectrum with little to no pre- or post-processing, seems likely to introduce oscillations into an already numerically delicate procedure.

\section{Results}\label{sec:results}
\begin{figure}[t!]
    \centering
    \includegraphics[width=1.0\linewidth]{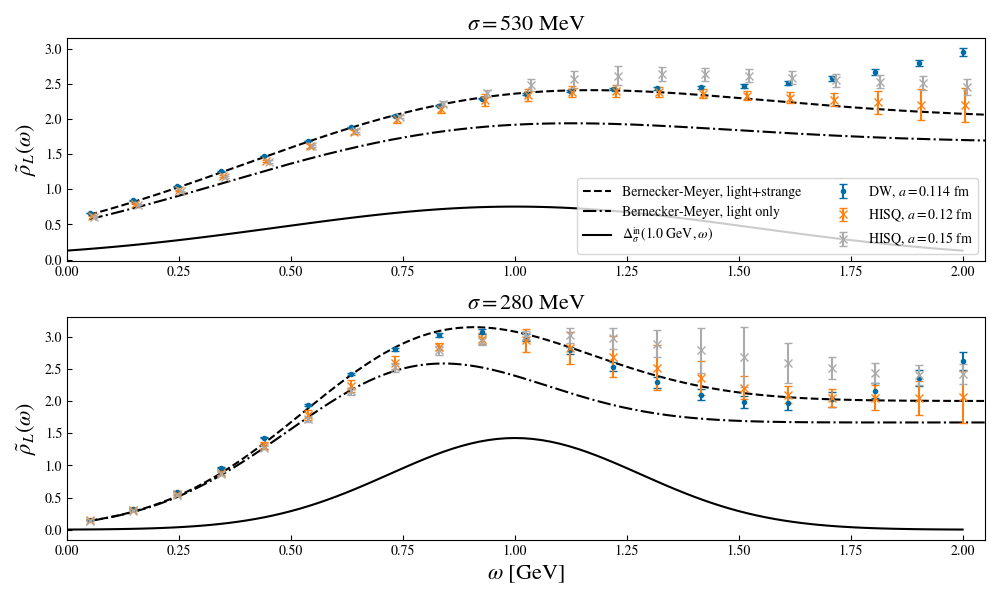}
    \caption{Spectral reconstructions of the domain wall and staggered datasets using the HLT method. $\Delta^\text{in}_\sigma$ is chosen to be a Gaussian with a smearing width of $\sigma$ (see Eq. \eqref{eq:gaussian}), and the basis functions defined in \cref{eq:basis_functions}.
    }
    \label{fig:reconstruction}
\end{figure}
Figure \ref{fig:reconstruction} shows reconstructions of the smeared $R$ ratio using the HLT algorithm applied to vector-vector correlation functions, using the analysis method described in \cref{sec:OSS} to handle the effects of opposite-parity states.
Results are shown for both the domain-wall and staggered datasets with Gaussian smearing widths $\sigma \in \{530, 280\}~{\rm MeV}$.
For comparison, the black dashed and dashed-dotted lines show the Bernecker-Meyer parameterization~\cite{RR_param} of the $R$~ratio after convolution with the Gaussian smearing kernel, modified to include light and strange and only light active quark flavors, respectively.
The errors in the reconstructions correspond to jackknife statistical errors and systematic errors (as defined as in Ref.~\cite{ExtendedTwistedMassCollaborationETMC:2022sta}) summed in quadrature.
The domain-wall reconstruction, computed with both light and strange quark contributions, does not require the same modifications as staggered reconstructions (and which, incidentally, also contain only light-quark contributions).
To this end we have included the domain wall reconstruction in this proceedings for preliminary qualitative comparison.
More detailed comparisons will appear in our forthcoming publication.

For the staggered reconstructions, a trend towards the light-quark smeared parameterization is observed.
The lattice discretization artifacts are known to be large in this case and trend downward with decreasing spacing for the local current, at least for some range of Euclidian time (see Ref.~\cite{FermilabLatticeHPQCD:2023jof}).
These discretization effects may be partly responsible for the similar results in the staggered and domain-wall datasets, despite the different valence-quark content.
Both the domain wall and the staggered fermion reconstructions differ from the Bernecker-Meyer parameterization at higher $(\gtrsim 1.25\text{ GeV})$ energies.
This deviation may be due to a combination of lattice artifacts and finite-volume effects.
Analysis of these effects will be performed in our forthcoming publication.

\section{Summary and Future Work}\label{sec:summary}
We have described techniques for spectral analysis of staggered-quark correlation functions
We have also reported preliminary spectral reconstructions of staggered-quark and domain-wall-quark correlation functions, leading to results for the smeared $R$~ratio.

At low energies $(< 1\text{ GeV})$ the smeared finite-volume spectral functions show promising agreement with the Bernecker-Meyer parameterization~\cite{RR_param}.
At energies $> 1\text{ GeV}$ however we note a departure from the parameterization.
This could be from lattice artifacts, finite-volume effects, or spectral states not included in the parameterization.
We plan to quantify the size of these effects in a forthcoming publication.

\section*{Acknowledgements} TB and DS were partially supported by the US DOE Office of Science under grant DE-SC0010339. LJ was supported by the DOE Office of
Science Early Career Award DE-SC002114.
WJ is support in part by the US DOE Office of Science under grant Contract Numbers DE-SC0011090 and DE-SC0021006.
The research reported in this work made use of computing and long-term storage
facilities of the USQCD Collaboration, which are funded by the Office of Science
of the U.S. Department of Energy.
Fermilab is managed by Fermi Research Alliance, LLC (FRA), acting under Contract No.~DE- AC02-07CH11359.
The authors would like to thank Alessandro De Santis for useful discussions.

\bibliographystyle{JHEP}
\bibliography{refs}

\end{document}